\documentclass[letterpaper, 10 pt, conference]{ieeeconf}  

\IEEEoverridecommandlockouts                              

\usepackage{amsmath} 
\usepackage{amssymb}  
\usepackage{bm}
\usepackage{graphicx}
\usepackage{float}
\usepackage{cite}
\usepackage{url}
\usepackage{algorithm}
\usepackage{algorithmic}
\usepackage{xcolor}

\renewcommand{\arraystretch}{0.9}

\title{\LARGE \bf
Nonlinear Model Predictive Control for 3D Formation of Multirotor Micro Aerial Vehicles with Relative Sensing in Local Coordinates*
}

\author{I. Kagan Erunsal$^{1,2}$   Rodrigo Ventura$^{1}$   Alcherio Martinoli$^{2}$
\thanks{*This work was partially supported by the FCT grant [PD/BD/135151/2017], the FCT doctoral program RBCog and the FCT project [UID/EEA/50009/2013].}
\thanks{$^{1}$Institute for Systems and Robotics (ISR), IST, Lisbon, Portugal}%
\thanks{$^{2}$Distributed Intelligent Systems and Algorithms Laboratory (DISAL), EPFL, Lausanne, Switzerland}%
}

\begin{document}

\maketitle
\thispagestyle{empty}
\pagestyle{empty}

\begin{abstract}
\label{sec:Abstract}
The complex tasks such as surveillance, construction, search and rescue can benefit of the maneuverability of multirotor Micro Aerial Vehicles (MAVs) to obtain robust, cooperative system behavior and formation control is a prominent component of the these complex tasks. This work focuses on the problem of three-dimensional formation control of multirotor MAVs by using exclusively relative sensory information. It proposes a centralized Nonlinear Model Predictive Control (NMPC) approach in a leader-follower scheme. A realistic six degrees of freedom mathematical model of a multirotor MAVs is introduced and leveraged in the control laws. The formulation of the problem is performed based on NMPC and relative sensing framework with respect to local coordinate frames of the robots. This type of formulation makes the formation independent of the full knowledge of global or common reference frames and the utilization of expensive global localization sensors. Real-time Iteration (RTI) based solution to optimal control problem (OCP) is proposed by taking the novel formulation into account. An extensive scenario is designed to test and validate the strategy. Evaluation of the results suggests that satisfactory robust performance is achieved and maintained under model uncertainty and noise in local sensors and even in cases where the dynamics of the formation suddenly changes.
\end{abstract}

\section{Introduction}
\label{sec:Intro}
For the last two decades, the coordination and cooperation strategies of multirotor Micro Aerial Vehicles (MAVs) have been drawn considerable attention of academy and industry since the deployment of multiple vehicles reduces the risk of mission failures and offers additional advantages such as increased sensing coverage through parallelism \cite{de2012formation},\cite{monteriu2015nonlinear}. In order to perform high level cooperative missions, the formation control usually becomes an essential component and Model Predictive Control (MPC) is a promising method to carry out this task deliberately. Due to its architectural flexibility and systematic handling of the performance and constraints, it is getting more attention nowadays \cite{eren2017model}. Among MPC methods, especially Nonlinear Model Predictive Control (NMPC) is particularly suitable to control the robots whose fast dynamics are needed to be predicted by nonlinear models and constraints as in multirotor MAVs. Furthermore, in order to deploy highly autonomous multirotor MAVs in non-trivial environments, several researchers focus on elaborating local sensing in formation control and try to solve its limitations \cite{schiano2016}.

The intersection of multi-rotor MAVs, formation control and MPC is well studied in the literature. Van Parys and Pipeleers \cite{van2017distributed}, investigate the 2D formation of quadrotor formation flight by ADMM-based distributed NMPC strategy in simulation. Bemporad and Rocchi \cite{bemporad2011decentralized} propose a cascaded decentralized linear time-varying MPC for leader-follower type of quadrotor formation problem in simulation. Freddi et al. \cite{monteriu2015nonlinear} present a decentralized NMPC approach for UAVs moving in leader-follower structure by leveraging communication and collision avoidance in a simulative case study. Yuan et al. \cite{yuan2017outdoor} introduce a decentralized MPC strategy for the outdoor flocking of quadrotors by using absolute sensing in 2D space and by discussing communication issues in simulation and real experiments consisting of five drones. In addition, in \cite{saska2014motion}, Saska et al. present both centralized motion planning and control methodology for the formation of MAVs with MPC in leader-follower scheme considering fault recovery and obstacle avoidance in simulation. Huck et al. \cite{huck2014rcopterx} investigate a distributed leader-follower linear MPC approach for the maintenance of formation shape as well as the reference tracking of whole formation and validated it with experiments consisting of three coaxial helicopters.

Relative sensing in formation control, without employing MPC framework, is also studied by several researchers. While Schiano et al. \cite{schiano2016} propose a rigidity-based bearing formation controller for a group of quadrotors, Franchi et al. \cite{franchi2016decentralized} present a decentralized control approach based upon local sensing for the encirclement problem in 3D. Finally, although Gowal and Martinoli \cite{gowal2012real} apply real-time receding-horizon optimization of trajectories that guarantee the rendezvous for only mobile robots, they presented a novel approach including relative sensing in local coordinate frames of the robots. For interested readers, other references include \cite{hafez2015solving},
\cite{fukushima2005distributed}, \cite{yamamato2017exp}, \cite{kuwata2007distributed}. Considering the presented literature, to the best of our knowledge, it can be concluded that there is no paper that formulates the formation control problem of multi-rotor MAVs with respect to local coordinate frames of the individual robots and elaborating relative sensing in a MPC framework. In this kind of approach, the formation is not dependent on the full knowledge of global or common coordinate frame and the sensory information coming from expensive global localization sensors, which might be a significant issue, for instance, indoor.

This paper aims the mentioned gap and presents a centralized NMPC strategy for the three-dimensional leader-follower formation control of multirotor MAVs. The first contribution is the novel formulation of the problem based-on exclusively relative sensing with respect to local coordinate frames of robots in an MPC framework. The most important problem here is how to select the model, inertial and body coordinate frames such that full state predictions are obtained independent of global variables. Furthermore, in this formulation, with the appropriate definition of sensing graph and controlled state variables, not only a rigid formation obtained but also the field of view restrictions (FOV) of relative sensors are satisfied automatically. The second contribution is to obtain the solution to optimal control problem (OCP) by employing modified Real-time Iteration (RTI) scheme to comply with real-time requirements and to have consistent warm-start of initial guess considering the formulation. The robustness of the method is shown in a challenging simulative experiment, including trajectory tracking of the group as a disturbance source, under noise and model uncertainties consisting of three vehicles. In every stage of the proposed method, real application aspects such as the selection of sensors, the compatibility of the strategy with working environment, solver performance evaluation etc. are taken into consideration.
  
The paper is organized as follows. In Section \ref{sec:Modeling}, the preliminary definitions and a realistic mathematical model of a multirotor MAV are presented. Section \ref{sec:ProbDef} gives a quick summary of the problem definition and assumptions. After Section \ref{sec:MPC} introduces and elaborates a centralized NMPC strategy, Section \ref{sec:SimRes} focuses on the implementation details of an extensive simulation and the results are presented and discussed. The conclusion and future work are finally provided in Section \ref{sec:Conc}.

\section{Basic Notations and Modeling}
\label{sec:Modeling}
In this section, the basic notations required for the rest of paper are introduced and the Six Degrees of Freedom (DOF) nonlinear mathematical model of an individual multirotor, in our case a quadrotor, is developed. The following notation for vectors and rotation matrices will be adopted assuming that $ \{a\}, \{b\}, \{c\} $ are the coordinate frames and $ x $ is the name of the vector:

\begin{description}
	\item[$ \bm{x}_{a/b}^{c}: $] Vector quantity $ x $ of origin of $ \{a\} $ with respect to $ \{b\} $ expressed in $ \{c\} $
	\item[$ \bm{R}^{b}_{a}: $] Rotation matrix $ R $ that transforms a vector expression from $ \{a\} $ to $ \{b\} $
\end{description}

\begin{figure}
	\centering
	\includegraphics[scale=0.6]{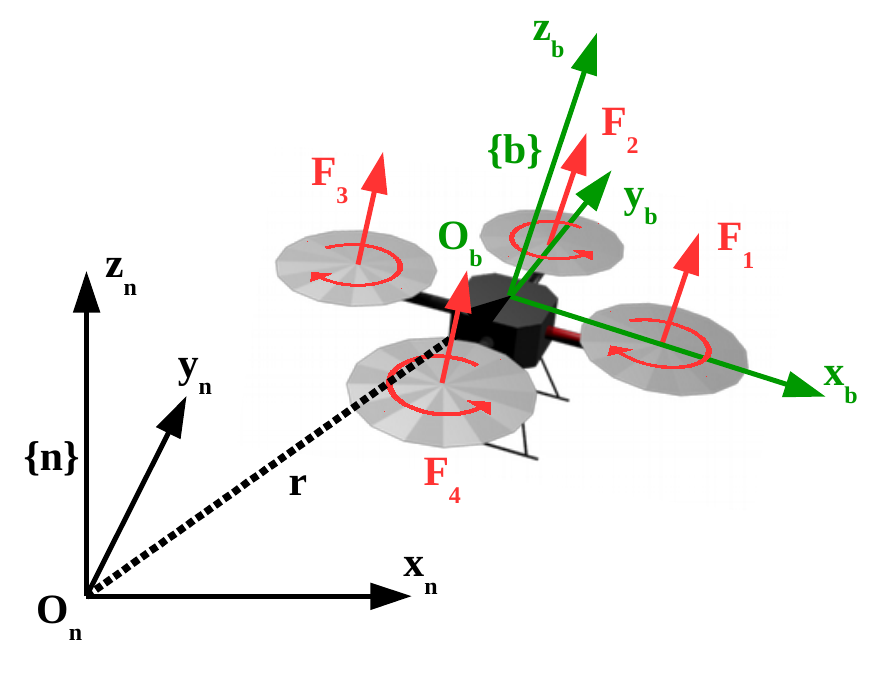}
	\caption{Earth and body coordinate frames of a quadrotor with its applied thrusts as a function of the rotational direction and the corresponding propeller.}
	\label{fig:CFs}
\end{figure}

For the mathematical modeling, there are two important coordinate frames for an individual quadrotor, namely earth-fixed inertial $ \{n\} $ and body-fixed $ \{b\} $ reference frames. These are illustrated in Fig. \ref{fig:CFs} along with the generated forces.

The rotation matrix between these two coordinate systems is given by  $ \bm{R}^{n}_{b} $ and it uses Z-Y-X Euler angles convention. The state column vector of the quadrotor is defined as in \eqref{eq:StateDef}.

\begin{equation}
	\label{eq:StateDef}
	\bm{state} = [{\bm{x}^n_{b/n}}^\top\ {\bm{v}^n_{b/n}}^\top\ {\bm{t}^n_{b/n}}^\top\ {\bm{w}^b_{b/n}}^\top]^\top
\end{equation}

\noindent where $ \bm{x} $ and  $ \bm{v} $ represent the position and linear velocity of $ \{b\} $ with respect to $ \{n\} $ expressed in $ \{n\} $,  $ \bm{t} $ is the Euler angles and $ \bm{w} $ is the angular velocities of $ \{b\} $ with respect to $ \{n\} $ expressed in $ \{b\} $. Based on this notation, rigid body dynamics of the airframe can be written by following four vector differential equations \eqref{eq:Dynamics1}-\eqref{eq:Dynamics4} \cite{mahony2012multirotor}.

\begin{eqnarray}
\label{eq:Dynamics1}
\dot{\bm{x}}^n_{b/n} &=& \bm{v}^n_{b/n} \\
\label{eq:Dynamics2}
m\dot{\bm{v}}^n_{b/n} &=& m_b\bm{g} + \bm{R}^{n}_{b} \bm{F}^{b}_{b/n} \\
\label{eq:Dynamics3}
\dot{\bm{t}}^n_{b/n} &=& \bm{T}^{n}_{b} \bm{w}^b_{b/n} \\
\label{eq:Dynamics4}
\bm{I}_b \dot{\bm{w}}^b_{b/n} &=& \bm{\tau}^{b}_{b/n} -  \bm{w}^b_{b/n} \times \bm{I}_b \bm{w}^b_{b/n}
\end{eqnarray}

\noindent where $ m_b $ denotes the mass of the airframe, $ \bm{g} $ represents the gravitational acceleration vector, $ \bm{T}^{n}_{b} $ is the angular transformation matrix, $ \bm{I}_b $ is the inertia of the vehicle with respect to its center of mass. Furthermore, due to dominant aerodynamic effects \cite{mahony2012multirotor}, there is a relation between the generalized forces and the propeller speeds as can be seen from \eqref{eq:Aerodynamics}.

\begin{equation}
\label{eq:Aerodynamics}
\left[\begin{array}{cccc}
\bm{F}^{b}_{b/ni,3} \\ \bm{\tau}^{b}_{b/n,1} \\ \bm{\tau}^{b}_{b/n,2} \\ \bm{\tau}^{b}_{b/n,3}
\end{array}\right] = \left[\begin{array}{cccc}
c_T & c_T & c_T & c_T \\
0 & d c_T & 0 & - d c_T \\
- d c_T & 0 &  d c_T  & 0 \\
c_Q & -c_Q & c_Q & -c_Q\end{array}\right]\left[\begin{array}{cccc}
\bm{\omega}_1^2 \\ \bm{\omega}_2^2  \\ \bm{\omega}_3^2  \\ \bm{\omega}_4^2
\end{array}\right]
\end{equation}

\noindent where $ c_T $ and $ c_Q $ denote the thrust and torque coefficients respectively, $ d $ represents the distance between a propeller center and the center of mass of the vehicle and $ \bm{\omega} $ represents the propeller speeds. The numbers in subscripts illustrate the element selection of the vectors.

\section{Problem Definition}
\label{sec:ProbDef}
\begin{figure}[!]
	\centering
	\includegraphics[scale=0.55]{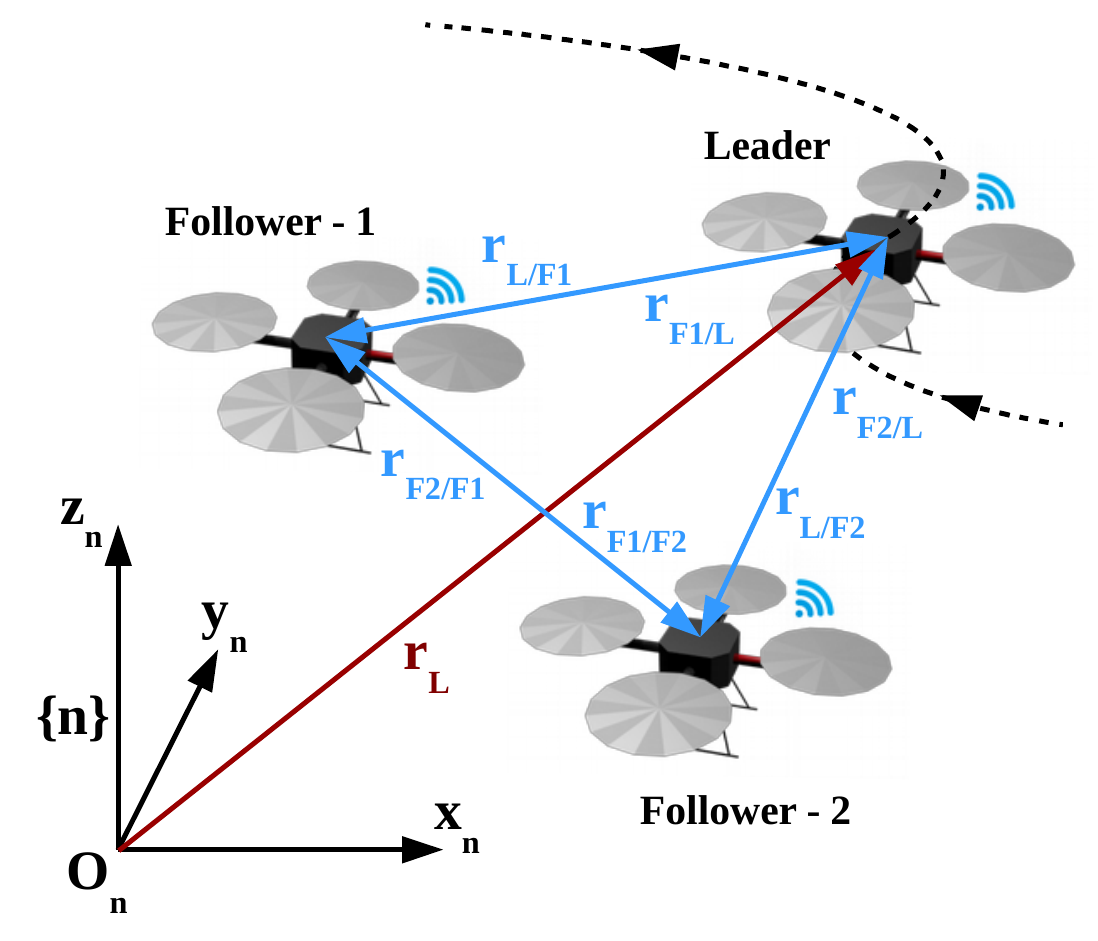}
	\caption{Formation control problem with one leader and two followers, the inter-vehicle positions and trajectory are also shown.}
	\label{fig:FormCont}
\end{figure}

This part focuses on the definition and categorization of the specific formation control problem. In this work, we are interested in the agents endowed exclusively with an on-board, limited FOV, 3D, relative localization system for measuring inter-vehicle positions in terms of relative range and bearing information. A concrete implementation of a similar sensor can be found in \cite{dias2016}. Furthermore, all agents are equipped with an optic flow-sonar sensor couple to obtain linear velocities and an IMU to acquire linear accelerations, rotational velocities and absolute roll and pitch information. Magnetometer cannot be accurately employed in the problem because it is assumed that the vehicles operate inside of a building. This implies that the absolute yaw information can not be obtained. There is only one leader in the team equipped with global sensors or a localization unit similar to one in \cite{Blosch2010} that obtains global information for position and velocity tracking. This is a logical assumption because indoor global localization sensors are expensive and require high onboard computation capacity. Note that trajectory tracking activity can be seen as both a disturbance to formation control and a secondary objective needs to be carried out along with formation. It is assumed that obstacles do not exist along the path. Throughout the scenario, while the leader's missions are both to maintain the formation and to follow a trajectory, the followers are responsible only for maintaining the formation. There is one central computation unit on the leader that executes the centralized NMPC approach and communicates with all followers to obtain their local sensor information and deliver propeller speed commands to them. Since only local sensory information and propeller speeds, not propagation of states or inputs, communicated, it is assumed that channel is not fully loaded. As a result, packet losses and delays do not exist in the channel. However, noise in all onboard sensors and uncertainties in modeling exist. This scenario is depicted in Fig. \ref{fig:FormCont}.

\section{Model Predictive Approach}
\label{sec:MPC}
Model Predictive Control (MPC) is an advanced control framework that had a large impact on industrial and academic control communities. This success is due to the fact that MPC is able to simultaneously handle constraints, design specifications and optimize the performance in a systematic manner \cite{eren2017model}. To achieve this outcome, MPC optimizes the predicted system behavior for a given horizon at the current time instance and finds the optimal control actions to be carried out within the horizon. Then, the chosen control actions are applied to the system and the procedure continues throughout the process. The applicability of MPC to very diverse system models and architectures makes this strategy even more attractive nowadays. However, MPC requires an accurate prediction of the system model and involves a computationally expensive procedure. These aspects make real-times applications much more difficult. Nevertheless, the considerable progress in the embedded computing capabilities and the significant advances in the dedicated numerical solvers mitigate the impact of the large computational requirements. If the model of the system that will be utilized in the MPC formulation or the constraints are nonlinear, NMPC is obtained. NMPC is preferred when the nonlinear models represent the system dynamics more accurately and this is the case for multi-rotor MAVs that carry-out agile maneuvers.

The OCP solved in each time instance for NMPC is shown in \eqref{eq:OCP} for a generic case \cite{acadowebsite2018}.

\begin{equation}
\label{eq:OCP}
\begin{aligned}
& \underset{\begin{subarray}{c}\bm{x}_0,...,\bm{x}_N \\
	\bm{u}_0,...\bm{u}_{N-1} \end{subarray}}{\text{minimize}}
& & \sum\limits_{k=0}^{N-1}c_k(\bm{x}_k,\bm{u}_k,\bm{r}_k)+c_{N}(\bm{x}_N,\bm{u}_N,\bm{r}_N) \\
&  \text{subject to}
& & \bm{x}_{k+1} = f_k(\bm{x}_k,\bm{u}_k), \quad k=0,..,N-1 \\
&&& g_k(\bm{x}_k,\bm{u}_k) = 0, \quad k=0,..,N-1 \\
&&& h_k(\bm{x}_k,\bm{u}_k) \leq 0, \quad k=0,..,N-1 \\
&&& g_N(\bm{x}_N) = 0
\end{aligned}
\end{equation}
 
\noindent where $ N $ is the selected horizon length, $ \bm{x}_0,...,\bm{x}_N $, $ \bm{u}_0,...,\bm{u}_N $ and $ \bm{r}_0,...,\bm{r}_N $ represent the evolution of the state, the input and the reference respectively, $ c_k $'s are the stage cost functions, $ c_N $ is the terminal cost function,  $ f_k $'s are used to express nonlinear dynamic equality constraints, $ g_k $'s and $ h_k $'s denote other, possibly nonlinear, equality and inequality constraints respectively and $ g_N $ stands for the terminal constraints. It should be noted that the dynamic equality constraints, i.e. equations of motion, given here are in discrete time, although the continuous time dynamics has been presented in Section \ref{sec:Modeling}. The transition is performed by employing a suitable discretization method such as Euler discretization and Runge-Kutta methods etc. The overall optimization problem is needed to be solved with respect to state and input sequence, considering that initial conditions are the current state and inputs, in each time instance for a given horizon. For large scale and complex problems, this is usually performed with numerical methods. Among the optimal and feasible solution sequence obtained for inputs, the one corresponds to initial instance are selected and applied to the plant. This procedure, feedback - optimization - application of inputs order, continues for each sampling instance until a desired objective reached.

In our problem, since the robots do not have the full knowledge of global or a common reference frames, for each horizon of the MPC, three new coordinate frames need to be defined to express the $c_k, c_N, f_k, g_k$ and $h_k$ uniquely:

\begin{itemize}
	\item \textbf{MPC body frame:} This frame is anchored to the body throughout the MPC horizon and it will be named as $ \{mb\} $. In real time, it coincides with the body-fixed coordinate frame $ \{b\} $.
	\item \textbf{Inertial MPC frame:} The position of this frame is fixed to the starting position of $ \{mb\} $ in the horizon. It has the same roll and pitch angle with $ \{n\} $  and has fixed yaw angle whose value is the same as the yaw angle of $ \{mb\} $ at the start of horizon. It will be represented by $ \{mi\} $. This frame is selected as inertial due to the fact that it is the simplest inertial frame that does not require absolute yaw information. Note that this frame is inertial in a MPC horizon and moves with non-zero acceleration in real time. 
	\item \textbf{MPC control frame:} The position is anchored to $ \{mb\} $, it has the same roll and pitch angle with $ \{n\} $ and it has the common yaw angle with $ \{mb\} $ during horizon. It will be named as $ \{mc\} $.  It is important to construct this frame for the formation control due to the fact that the roll and the pitch angles of a quadrotor cannot be controlled directly during the motion. The reference frame that the controlled variables are defined should be independent of this type of uncontrolled states. 	
\end{itemize}

\begin{figure}[!]
	\centering
	\includegraphics[scale=0.8]{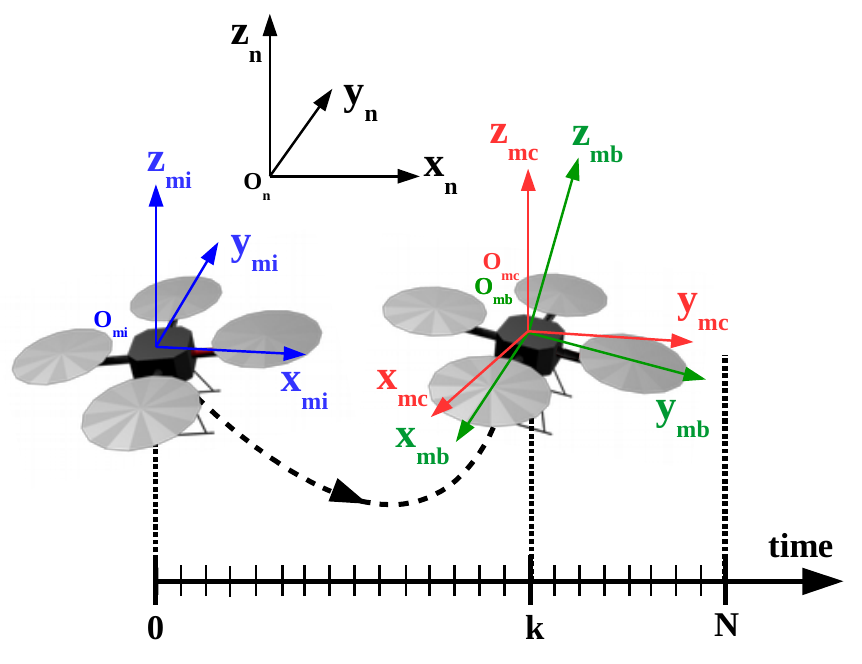}
	\caption {MPC coordinate frames in a horizon: $ \{mi\} $ (blue), $ \{mb\} $ (green), $ \{mc\} $ (red).}
	\label{fig:MPC_CoorFrames}
\end{figure}

\noindent Note that, all frames are local to robots and not common among them.  These frames including the earth-fixed reference frame $ \{n\} $ are illustrated in Fig. \ref{fig:MPC_CoorFrames}.

To formulate the nonlinear dynamic equality constraints, the mathematical model of the system should be rewritten considering these MPC frames. Initially, the states are redefined with respect to $ \{mi\} $ and $ \{mb\} $  as in \eqref{eq:MPC_StateDef}.

\begin{equation}
\label{eq:MPC_StateDef}
\bm{s}_{m} = [{\bm{x}^{mi}_{mb/mi}}^\top\ {\bm{v}^{mi}_{mb/mi}}^\top\ {\bm{t}^{mi}_{mb/mi}}^\top\ {\bm{w}^{mb}_{mb/mi}}^\top]^\top
\end{equation}

\noindent Then, the system model is reformulated for both the leader and the followers as in \eqref{eq:MPC_Dynamics1} - \eqref{eq:MPC_Dynamics4}.

\begin{align}
\label{eq:MPC_Dynamics1}
\dot{\bm{x}}^{mi}_{mb/mi} &= \bm{v}^{mi}_{mb/mi} \\
\label{eq:MPC_Dynamics2}
m_{bm}\dot{\bm{v}}^{mi}_{mb/mi} &= \bm{R}^{mi}_{n}m_{bm}\bm{g} + \bm{R}^{mi}_{mb} \bm{F}^{mb}_{mb/mi} \\
\label{eq:MPC_Dynamics3}
\dot{\bm{t}}^{mi}_{mb/mi} &= \bm{T}^{mi}_{mb} \bm{w}^{mb}_{mb/mi} \\
\label{eq:MPC_Dynamics4}
\bm{I}_{bm} \dot{\bm{w}}^{mb}_{mb/mi} &= \bm{\tau}^{mb}_{mb/mi} -  \bm{w}^{mb}_{mb/mi} \times \bm{I}_{bm} \bm{w}^{mb}_{mb/mi}
\end{align}
\noindent Note that the rotation matrix $ \bm{R}^{mi}_{n} $ that transforms gravitational acceleration vector to $ \{mi\} $ does not change the value of $ \bm{g} $ after transformation due to selection of $ \{mi\} $. Furthermore, in order to predict the future states from \eqref{eq:MPC_Dynamics1} - \eqref{eq:MPC_Dynamics4}, no global information is needed except roll and pitch Euler angles which is included in $\bm{R}^{mi}_{mb}$ and are provided by IMU. Additionally, since the leader is supposed to perform target tracking, the position, velocity and the Euler angles of leader should be defined with respect to $ \{n\} $ in terms of these new MPC states. This formulation can be found in \eqref{eq:Leader_States1} - \eqref{eq:Leader_States3}.

\begin{align}
\label{eq:Leader_States1}
\bm{x}^{n}_{mb/n} &=  \bm{R}^{n}_{mi}\bm{x}^{mi}_{mb/mi}+\bm{x}^{n}_{mi/n} \\
\label{eq:Leader_States2}
\bm{v}^{n}_{mb/n} &= \bm{R}^{n}_{mi}\bm{v}^{mi}_{mb/mi} \\
\label{eq:Leader_States3}
\bm{t}^{n}_{mb/n} &= \bm{t}^{mi}_{mb/mi} + \bm{t}^{n}_{mi/n}
\end{align}

\begin{figure}[!]
	\centering
	\includegraphics[scale=0.78]{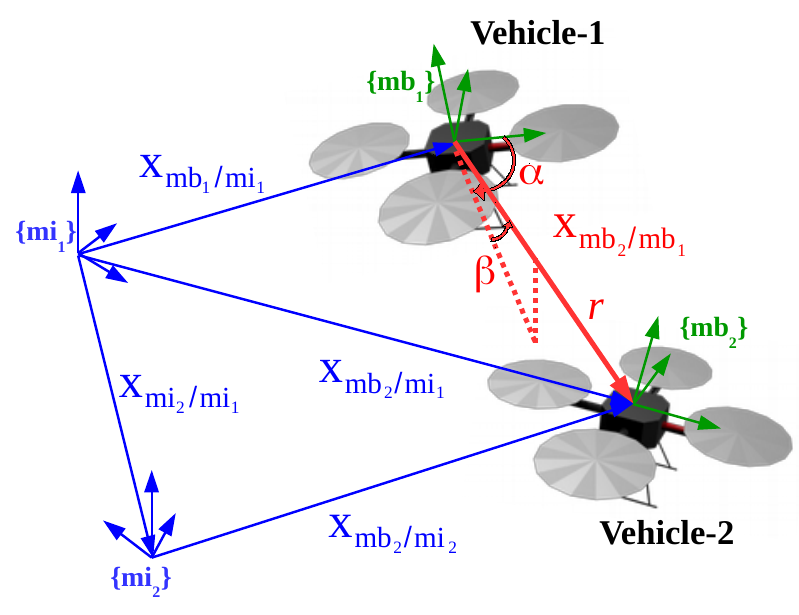}
	\caption {Graphical representation of the relative displacement between two neighbor vehicles, the range and the bearing are also shown.}
	\label{fig:RPs}
\end{figure}

\noindent In order to obtain the controlled variables, it is important to formulate them in terms of MPC states as well. As indicated in Fig. \ref{fig:RPs}, the relative displacement vector between any two vehicle, vehicle 1 and vehicle 2 in this case, is calculated as in \eqref{eq:RelPos1} - \eqref{eq:RelPos4}.

\begin{eqnarray}
\label{eq:RelPos1}
\bm{x}^{mi_1}_{mb_2/mi_2} & = &\bm{R}^{mi_1}_{mi_2}\bm{x}^{mi_2}_{mb_2/mi_2} \\
\label{eq:RelPos2}
\bm{x}^{mi_1}_{mb_2/mi_1} & = & \bm{x}^{mi_1}_{mb_2/mi_2} + \bm{x}^{mi_1}_{mi_2/mi_1} \\
\label{eq:RelPos3}
\bm{x}^{mi_1}_{mb_2/mb_1} & = & \bm{x}^{mi_1}_{mb_2/mi_1} - \bm{x}^{mi_1}_{mb_1/mi_1} \\
\label{eq:RelPos4}
\bm{x}^{mc_1}_{mb_2/mb_1} & = &\bm{R}^{mc_1}_{mi_1}\bm{x}^{mi_1}_{mb_2/mb_1}
\end{eqnarray}

\noindent Note the rotation matrix $ \bm{R}^{mi_1}_{mi_2} $ between two inertial MPC frames of the vehicles. Since two reference frames are different from each other only in yaw angle orientation-wise, the relative yaw angle estimation is necessary to calculate this matrix. As indicated previously, there is no absolute yaw measurement for the followers. Since the control architecture is centralized and has access to both of the local measurements, the range and the bearing feedback of the vehicles can be utilized to generate this information. Assume that $ \bm{x}^{mb_1}_{mb_2/mb_1} $  and $ \bm{x}^{mb_2}_{mb_1/mb_2} $ are the relative displacement vectors of the vehicles expressed in their own body-fixed frames. These vectors can also be expressed in terms of the azimuth $ \alpha $, the elevation $ \beta $  and the range $ r $  information of the relative sensors as in \eqref{eq:CameraRange1} and \eqref{eq:CameraRange2} and they are depicted in Fig. \ref{fig:RPs}.  

\begin{align}
\nonumber 
\bm{x}^{mb_1}_{mb_2/mb_1} =& [r_{2/1}\cos(\beta_1)\cos(\alpha_1) \\
& \label{eq:CameraRange1}
r_{2/1}\cos(\beta_1)\sin(\alpha_1)\ r_{2/1}\sin(\beta_1)]^\top \\ \nonumber
\bm{x}^{mb_2}_{mb_1/mb_2} =& [r_{1/2}\cos(\beta_2)\cos(\alpha_2) \\&\label{eq:CameraRange2}   
r_{1/2}\cos(\beta_2)\sin(\alpha_2)\ r_{1/2}\sin(\beta_2)]^\top	
\end{align}

\noindent Then, a relation between the relative displacements can be written as in \eqref{eq:YawEst1} by expressing them in $ \{n\} $.

\begin{equation}
\label{eq:YawEst1}
\bm{R}^{n}_{mb_1}\bm{x}^{mb_1}_{mb_2/mb_1}+\bm{R}^{n}_{mb_2}\bm{x}^{mb_2}_{mb_1/mb_2} = 0
\end{equation}

\noindent Expanding the rotation matrices in terms of individual Euler angles gives the equation in \eqref{eq:YawEst2}.

\begin{multline}
\label{eq:YawEst2}
\bm{R}^{n}_{z,mb_1}\bm{R}^{n}_{y,mb_1}\bm{R}^{n}_{x,mb_1}\bm{x}^{mb_1}_{mb_2/mb_1}+ \\\bm{R}^{n}_{z,mb_2}\bm{R}^{n}_{y,mb_2}\bm{R}^{n}_{x,mb_2}\bm{x}^{mb_2}_{mb_1/mb_2} = 0
\end{multline}

\noindent Since the absolute roll and pitch information are known, corresponding rotation matrices are also known. Rearranging the equation and separating the known terms as vectors $ a $ and $ b $ result in a new expression \eqref{eq:YawEst3}.

\begin{equation}
\label{eq:YawEst3}
{\bm{R}^{n}_{z,mb_2}}^{-1}\bm{R}^{n}_{z,mb_1}b - a = 0
\end{equation}

\noindent which gives a simpler expression in terms of relative yaw angles as in (\ref{eq:YawEst4}).

\begin{equation}
\label{eq:YawEst4}
\bm{R}^{n}_{z}(\psi_1-\psi_2)b - a = 0
\end{equation}

\noindent where $\psi$ represents absolute yaw angle.  This equation can be solved by considering it as an optimization problem and the solution yields the following expression \eqref{eq:YawEst5}.

\begin{multline}
\label{eq:YawEst5}
\psi_1-\psi_2 = {\rm atan2} (a_2b_1-a_1b_2, a_1b_1+a_2b_2)
\end{multline}

\noindent where subscripts for $ a $ and $ b $ show element selection. Now, everything is set up to write the cost function. The horizon objective function consist of three parts: formation, trajectory tracking and input related as in \eqref{eq:CostFunc1}. On the other hand, the terminal objective function includes two parts: formation and trajectory tracking related parts as in \eqref{eq:CostFunc2}.

\begin{align}
\nonumber c_k =& \sum\limits_{i=1}^{n_i}\sum\limits_{j=1}^{n_j} \| \widetilde{\bm{x}}^{mc_i}_{k,mb_j/mb_i}-\bm{x}^{mc_i}_{k,mb_j/mb_i} \|^2_{W^f_{kij}}+ \\ &\nonumber
\| \widetilde{\bm{x}}^{n_l}_{k,mb_l/n_l}-\bm{x}^{n_l}_{k,mb_l/n_l} \|^2_{W^x_k}+ \\ &\label{eq:CostFunc1}  \|\widetilde{\psi}^{n_l}_{k,mb_l/n_l}-\psi^{n_l}_{k,mb_l/n_l} \|^2_{W^t_k}+ \\ &\nonumber
\|\bm{R}^{mi}_{k,mb} \bm{F}^{mb}_{k,mb/mi}-g\|^2_{W^{F}_{kij}} \\ &\nonumber
\|\bm{\tau}^{mb}_{mb/mi}\|^2_{W^{\tau}_{kij}}\\
\nonumber c_N =& \sum\limits_{i=1}^{n_i}\sum\limits_{j=1}^{n_j} \| \widetilde{\bm{x}}^{mc_i}_{N,mb_j/mb_i}-\bm{x}^{mc_i}_{N,mb_j/mb_i} \|^2_{W^f_{Nij}}+\\& \label{eq:CostFunc2} 
\| \widetilde{\bm{x}}^{n_l}_{N,mb_l/n_l}-\bm{x}^{n_l}_{N,mb_l/n_l} \|^2_{W^x_N}+\\& \nonumber \|\widetilde{\psi}^{n_l}_{N,mb_l/n_l}-\psi^{n_l}_{N,mb_l/n_l} \|^2_{W^t_N}
\end{align}

\noindent where $ n_i $ represents the number of vehicles, $ n_j $ indicates the number of neighbors, $ W $s are the corresponding weight matrices and $ \sim $  denotes the references. Note that cost function includes the relative displacement errors in both ways, namely from agent $ i $ to $ j $ and from $ j $ to $ i $. This causes not only rigid formation configuration but also collision avoidance between followers. Furthermore, note that for input minimizing terms, non-gravitational inertial force and body fixed moments are minimized in order to obtain smooth output. Apart from the cost function, the constraints should also be specified. For this study, while $ h_k $'s, as in \eqref{eq:OCP}, are utilized to define state and input upper and lower bounds, $ g_k $'s represent the inter-stage algebraic equality constraints.

\begin{figure}[!]
	\centering
	\includegraphics[scale=0.4]{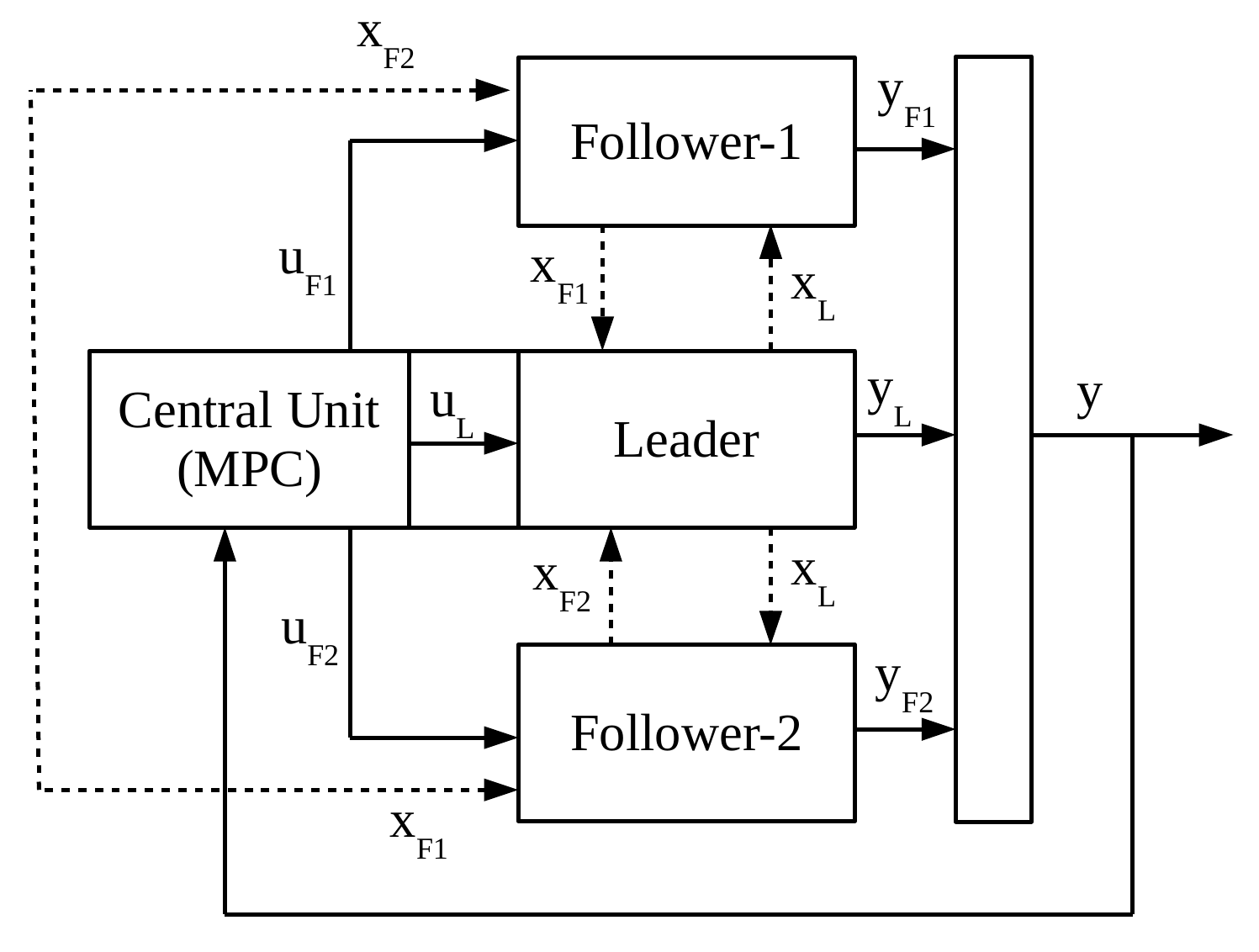}
	\caption {Centralized MPC structure in formation control.}
	\label{fig:ContStruct}
\end{figure}

At this point, it is worth mentioning the controller structure. As indicated previously, the control scheme is centralized and is illustrated in Fig. \ref{fig:ContStruct}. While the solid lines show the communication requirement between agents, the dashed lines indicate the possible effect of the state of one vehicle to another. In this figure, $ u, x, y $ represent the input, state and output (processed sensor information), respectively. Furthermore, communication-wise, the graph of the formation is a tree \cite{oh2015survey} and sensing-wise it is undirected and complete. Although centralized approaches suffer from poor scalability, they are well suited for small groups of UAVs. This undirectedness property of sensing graph and definition of controlled variables in local frames cause the followers to follow not only positions but also yaw angle of the leader. Lastly, the inclusion of the sensing/control branch between followers provide additional collision avoidance.

To solve (\ref{eq:OCP}), an efficient and fast solver generated by ACADO toolkit by  Houska et al. \cite{Houska2011a} is employed. This solver is selected due to the fact that it has open-source, user-friendly MATLAB interface, it can be easily modified, it has fast C code generation for real-time implementation. This solver utilizes the Real-time Iteration (RTI) scheme based on Gauss-Newton approach \cite{kamel2017linear}. First of all, the system dynamics and the other constraints are discretized over the horizon to obtain a structured Nonlinear Program (NLP). Then, Sequential Quadratic Programming (SQP) type of algorithm approximates the NLP by convex Quadratic Program (QP) sub-problems. After applying condensing, next step is to use a linear algebra solver such as qpOASES to obtain a solution for the sub-problems directly \cite{quirynen2015autogenerating}. Generally, the warm-start property of RTI scheme is applied to obtain better initial solutions since this scheme is rely on the fact that optimal solution to current iteration is very similar to previous one.

In this problem, the modified RTI scheme is employed and given in Algorithm \ref{alg:MRTI} for individual control iteration. Initially, before operation, the algorithm tries to obtain the best solution considering output KKT tolerance of a single SQP step, which assesses the quality of solution \cite{van2014real}. After it obtains a low KKT tolerance solution, in operation, it decides how many SQP steps to perform considering the computation time of solution, control sample time and desired KKT tolerance. The infeasible solutions are handled by using previous predicted solutions. For feasible solutions, the warm start of the next iteration is prepared by shifting the optimal solution in time and applying a transformation. This transformation is required due to the fact that all MPC frames (inertial, body and control) on which the problem formulation is carried out will translate and rotate small amount in the next iteration. Note that this transformation (rotation and translation) is known by the state predictions of the current iteration for the next step. 

\begin{algorithm}[!]
	\algsetup{linenosize=\scriptsize}
	\scriptsize
	\caption{Modified RTI scheme}
	\begin{algorithmic}
		\label{alg:MRTI}
		\STATE $X^g \leftarrow x^g$ \COMMENT{\textcolor{blue}{Initial guess of evolution of states}}
		\STATE $U^g \leftarrow u^g$ \COMMENT{\textcolor{blue}{Initial guess of evolution of inputs}}
		\STATE $X_0 \leftarrow x_0$ \COMMENT{\textcolor{blue}{Feedback obtained, initial condition}}
		\STATE $P_s \leftarrow p_s$ \COMMENT{\textcolor{blue}{Parameters and online data}}
		\WHILE{($t = 0 \enspace|\enspace t_{CPU} \leq t_s) \enspace\&\enspace KKT value \geq threshold$} \STATE \COMMENT{\textcolor{blue}{Initially iterate until obtain a low KKT tolerance result, for the next iterations check CPU time, control sample time and KKT tolerance to finish}}
		\STATE $(X^*,U^*) \leftarrow SQP\_step(X^g,U^g,X_0,P_s)$ \COMMENT{\textcolor{blue}{Obtain optimal states and inputs by SQP}}
		\IF{$(X^*,U^*$) is infeasable}
		\STATE $iflag \leftarrow true$ \COMMENT{\textcolor{blue}{Infeasibility flag}}
		\STATE \textbf{break}
		\ENDIF
		\STATE $(X^g,U^g) \leftarrow (X^*,U^*)$ \COMMENT{\textcolor{blue}{Warm start for next sub-iteration}}
		\ENDWHILE	
		\IF {$iflag = true$}
		\STATE \textbf{Apply} $U^g(1)$
		\STATE $(X^g,U^g) \leftarrow T(X^g_s,U^g_s)$ \COMMENT{\textcolor{blue}{Warm start for next iteration: Shifted and transformed version of the previous guess}}
		\ELSE
		\STATE \textbf{Apply} $U^*(1)$
		\STATE $(X^g,U^g) \leftarrow T(X^*_s,U^*_s)$ \COMMENT{\textcolor{blue}{Warm start for next iteration: Shifted and transformed version of the optimal solution}}		
		\ENDIF 
	\end{algorithmic}
\end{algorithm}

\section{Simulation and Results}
\label{sec:SimRes}
The mathematical modeling, controller implementation, solver generation and  simulation have been performed in MATLAB 2017a. The hardware utilized is Intel Core i7-7700HQ processor. One leader and two followers are employed. This scenario is designed in a way to include the following test cases: static formation maintenance, formation control in linear motion, in agile turning, in spiral motion and formation regeneration. The trajectories to be tracked are determined by considering the speed, acceleration and force limits except for the agile turning case. Note that the trajectory also consists of yaw references which might be an important aspect for quadrotors in formation. The properties of the designed trajectory are given as:

\begin{itemize}
	\item \textbf{(A) Static formation:} 1 sec. long formation maintenance.
	\item \textbf{(B) Formation control in linear motion:} Linear path tracking with the speed of 2 $m/s$, the yaw angle is constant.
	\item \textbf{(C) Formation control in agile turning:} 90 degrees turning during the linear trajectory tracking, the yaw angle tracks the change.
	\item \textbf{(D) Formation regeneration:} This is where the formation reference changes to produce a new shape.
	\item \textbf{(E) Formation control in spiral motion:} Spiral trajectory tracking at the speed of 1.9 m/s. The yaw angle tracks the trajectory as well.
\end{itemize}

Throughout the scenario, for all robots, optic flow sensor, IMU and relative localization sensor have inherited zero mean Gaussian noises. Furthermore, for the leader, the absolute localization unit also has zero mean Gaussian noise. The standard deviation of noises are determined from Pixhawk documentation, its peripherals and from the papers \cite{garcia2018SensorFusion} and \cite{honegger2013open}. Furthermore, the prediction model of NMPC has parameter uncertainty in the mass and inertia of quadrotors. \noindent The important parameters utilized in the simulation including noise and uncertainty values are provided in Table \ref{tab:FC_Params}.

\begin{table}[!]
	\scriptsize
	\renewcommand{\arraystretch}{1.0}
	\caption{Selected simulation parameters.}
	\label{tab:FC_Params}
	\centering
	\begin{tabular}{c||c||c}
		\hline
		\bfseries Parameter & \bfseries Value & \bfseries Unit \\ 
		\hline\hline
		Vehicle characteristic length & 0.18 & m \\
		\hline
		Maximum propeller speed & 6000 & rpm \\
		\hline
		Plant and control sampling time & 0.05 & s \\
		\hline
		Horizon length & 15 & - \\
		\hline
		Hessian approximation & Gauss-Newton & - \\
		\hline
		Condensing algorithm & Full Condensing & - \\
		\hline
		Discretization type & Multiple Shooting & - \\
		\hline
		Integrator type & INT-RK4 & - \\
		\hline
		Initial admissible KKT tolerance & $10^{-3}$ & - \\
		\hline
		Running admissible KKT tolerance & $10$ & - \\
		\hline
		Formation ref. F1 w.r.t. L & $[-1\ -0.5\ -0.5]^\top$ & m \\
		\hline
		Formation ref. F1 w.r.t. F2 & $[0\ -1\ 0]^\top$ & m \\
		\hline
		Formation ref. F2 w.r.t. L & $[-1\ 0.5\ -0.5]^\top$ & m \\
		\hline
		Updated formation ref. F1 w.r.t. L & $[-1\ -0.75\ -0.5]^\top$ & m \\
		\hline
		Updated formation ref. F1 w.r.t. F2 & $[0\ -1.5\ 0]^\top$ & m \\
		\hline
		Updated formation ref. F2 w.r.t. L & $[-1\ 0.75\ -0.5]^\top$ & m \\
		\hline
		Rel. weights of form., track., inputs & 10, 1, 1 & - \\
		\hline
		Std. dev. of abs. loc. noise & 0.02 & m \\
		\hline
		Std. dev. of optic flow noise & 0.25 & m/s \\
		\hline
		Std. dev. of IMU noise (Euler) & 0.005 & deg \\
		\hline
		Std. dev. of gyro noise & 3 & deg/s \\
		\hline
		Std. dev. of rel. loc. noise & 0.025 & m \\
		\hline
		Mass and inertia uncertainty (Std. dev.) & 1\%, 5\% & - \\
		\hline
	\end{tabular}
\end{table}

The important parameters such as horizon length, sampling time and weight matrices are determined by trial and error and considering the following aspects: Stability of the vehicles and formation, maximum CPU time required for the solver, control sample time, desired accuracy in error norms of formation and trajectory tracking.

\begin{figure}[!]
	\centering
	\includegraphics[scale=0.28]{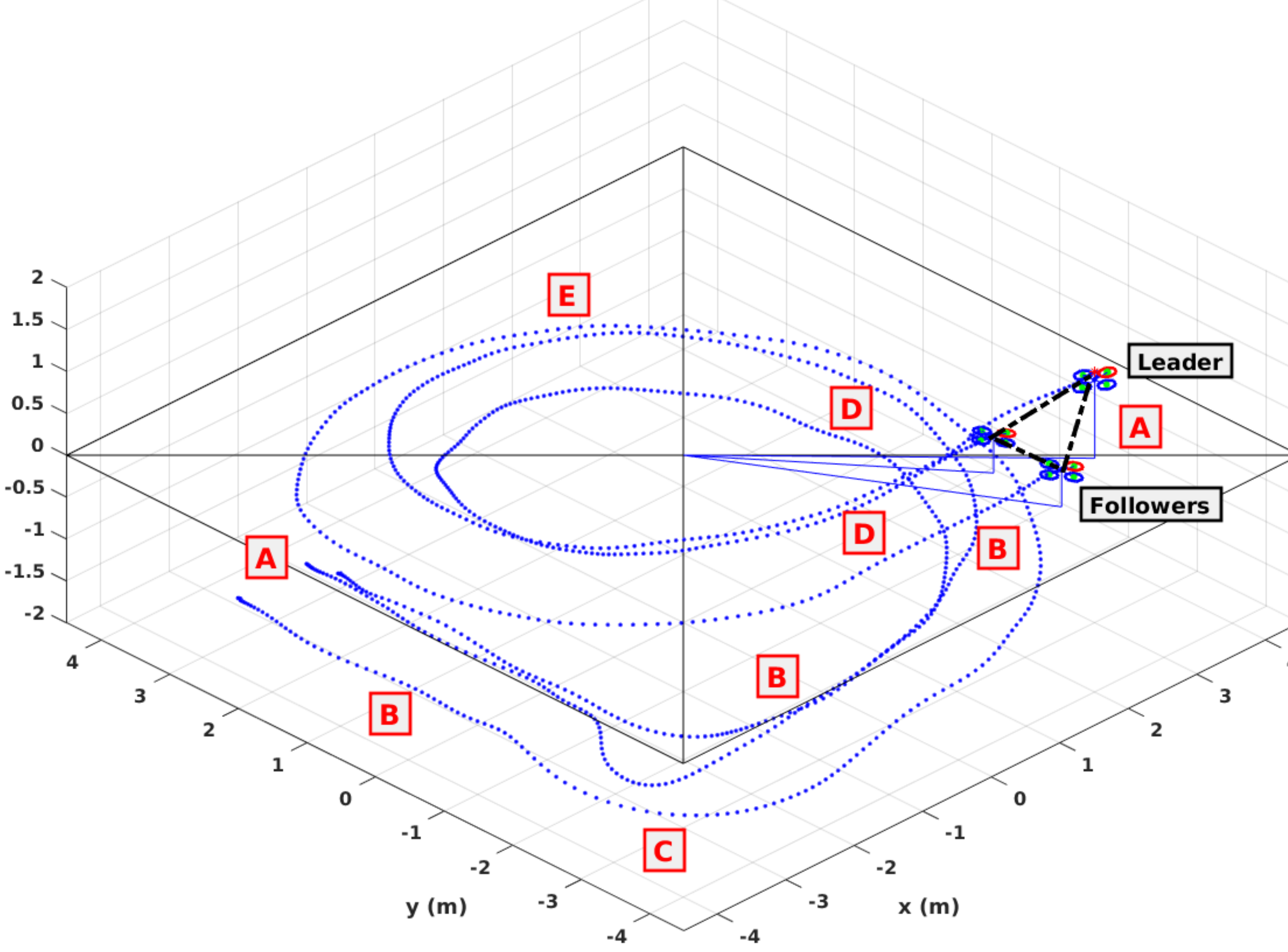}
	\caption {3D trajectory followed by the vehicles in different scenarios.}
	\label{fig:3DTraj}
\end{figure}

The 3D trajectories followed by the vehicles are illustrated in Fig. \ref{fig:3DTraj} the with corresponding letters. Although this figure only gives a general idea about the scenario and the trajectories, a careful examination shows that performance is quite satisfactory for all cases.

\begin{figure}[!]
	\centering
	\includegraphics[scale=0.30]{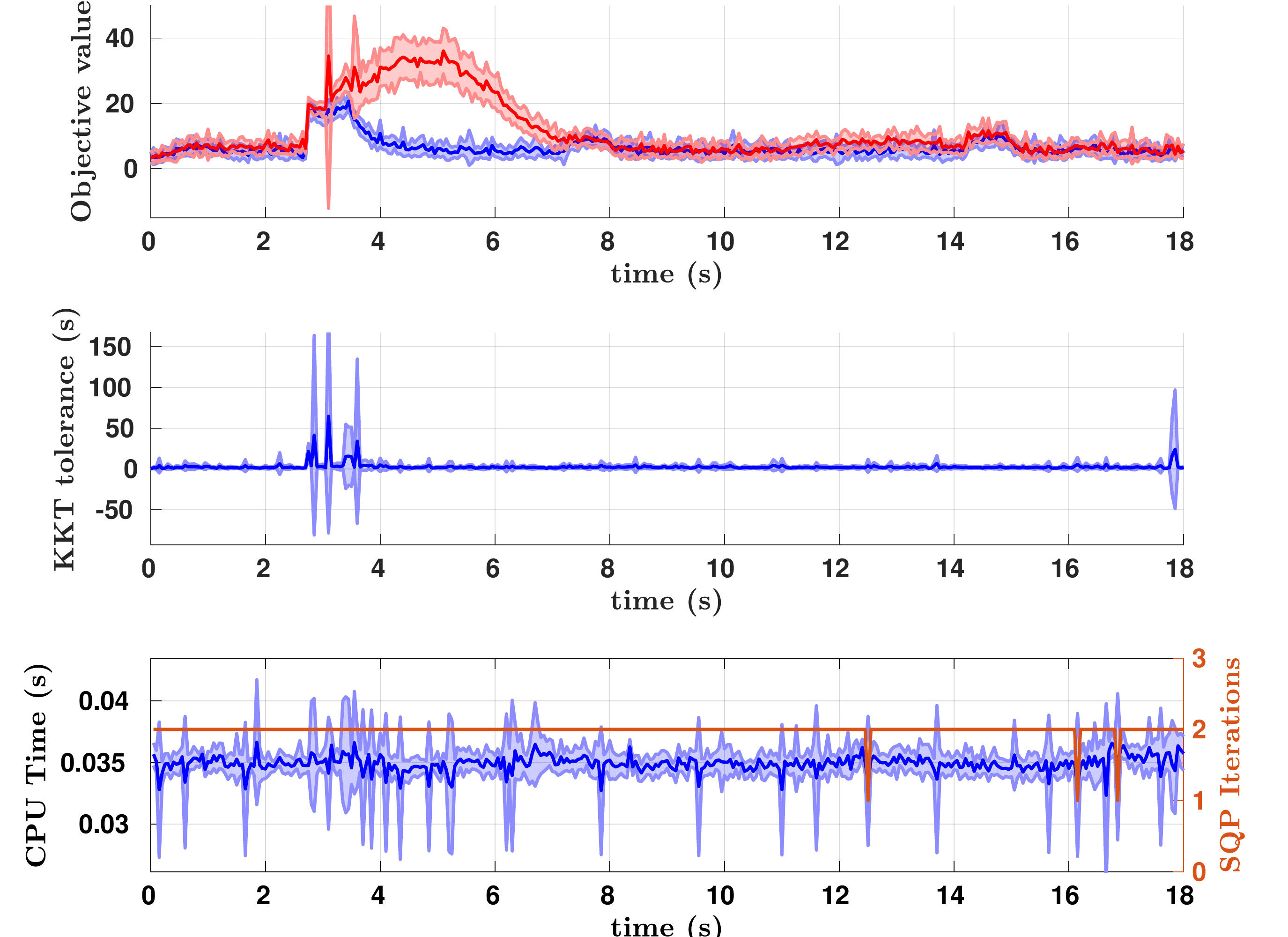}
	\caption {Performance measures of the solver: The mean and standard deviation of objective function, KKT tolerance, CPU time. Blue and orange: with modified RTI scheme, Red: without warm start and transformations.}
	\label{fig:Solver}
\end{figure}

The measures for solver performance throughout the scenario are given in Fig. \ref{fig:Solver} for 10 runs. As demonstrated, the objective function value has three peaks in agile turning and formation regeneration cases but all peaks diminish after sometime thanks to the predicted behavior of NMPC. Furthermore, the oscillating behavior of the cost function value is due to the noise in sensors and uncertainty in model parameters. Additionally, it can be observed that without applying modified RTI scheme, the objective function value is much higher especially in agile turning. Although they are not given here, the same performance degradation can be seen in the KKT tolerance and error norms. Considering the KKT values, it is observed that in average, RTI obtains a small KKT tolerance which means that SQP's converge to Nonlinear Program (NLP). Finally CPU time shows that it is always below control sample time, which is 0.05 \textit{seconds} and for the vast majority of each sample time, 2 SQP iterations are performed to reduce the tolerance.

\begin{figure}[t]
	\centering
	\includegraphics[scale=0.28]{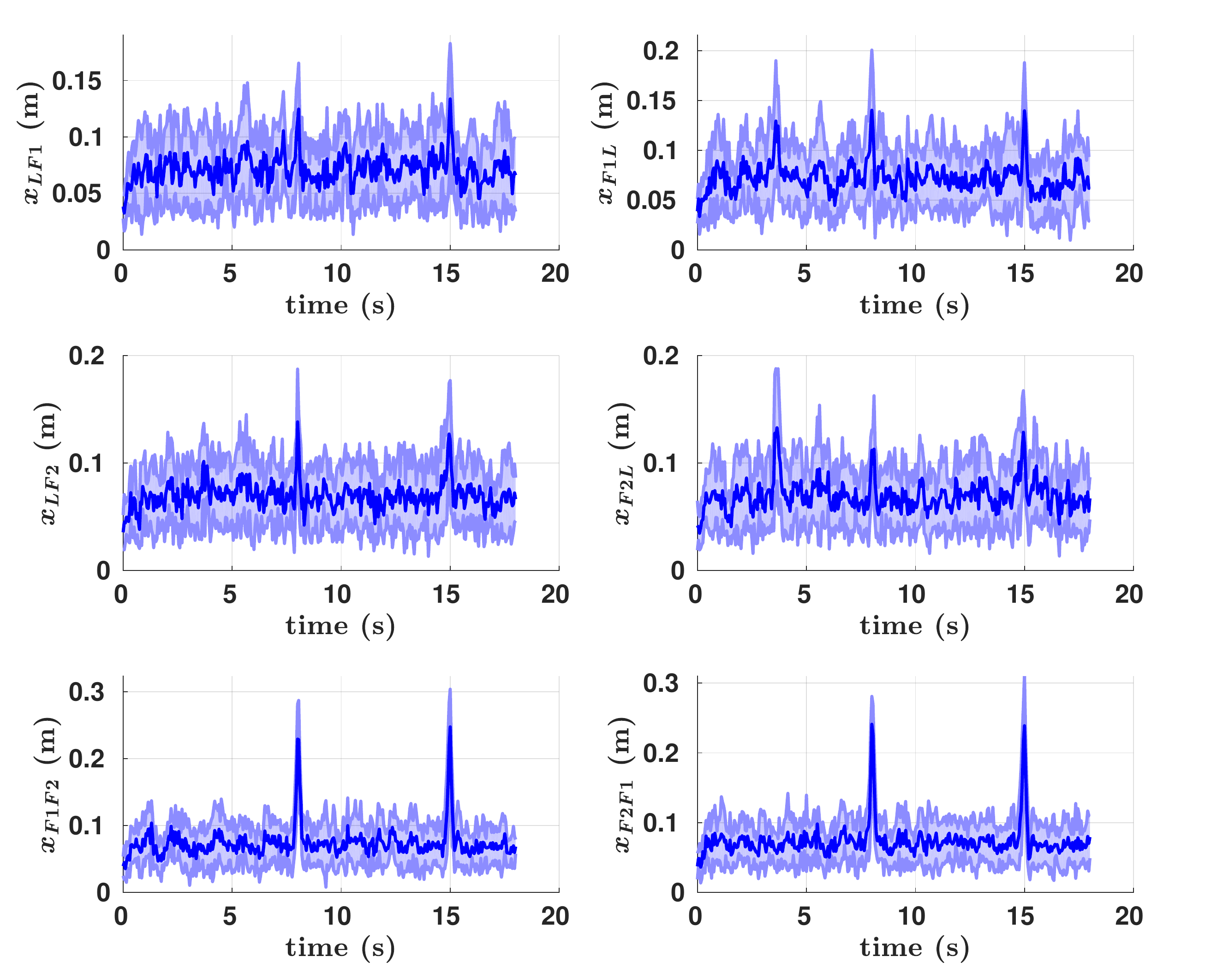}
	\caption {The mean and standard deviations of error norms of relative positions between robots: L: Leader, F1: Follower-1, F2: Follower-2}
	\label{fig:FormErrs}
\end{figure}

\begin{figure}[!]
	\centering
	\includegraphics[scale=0.24]{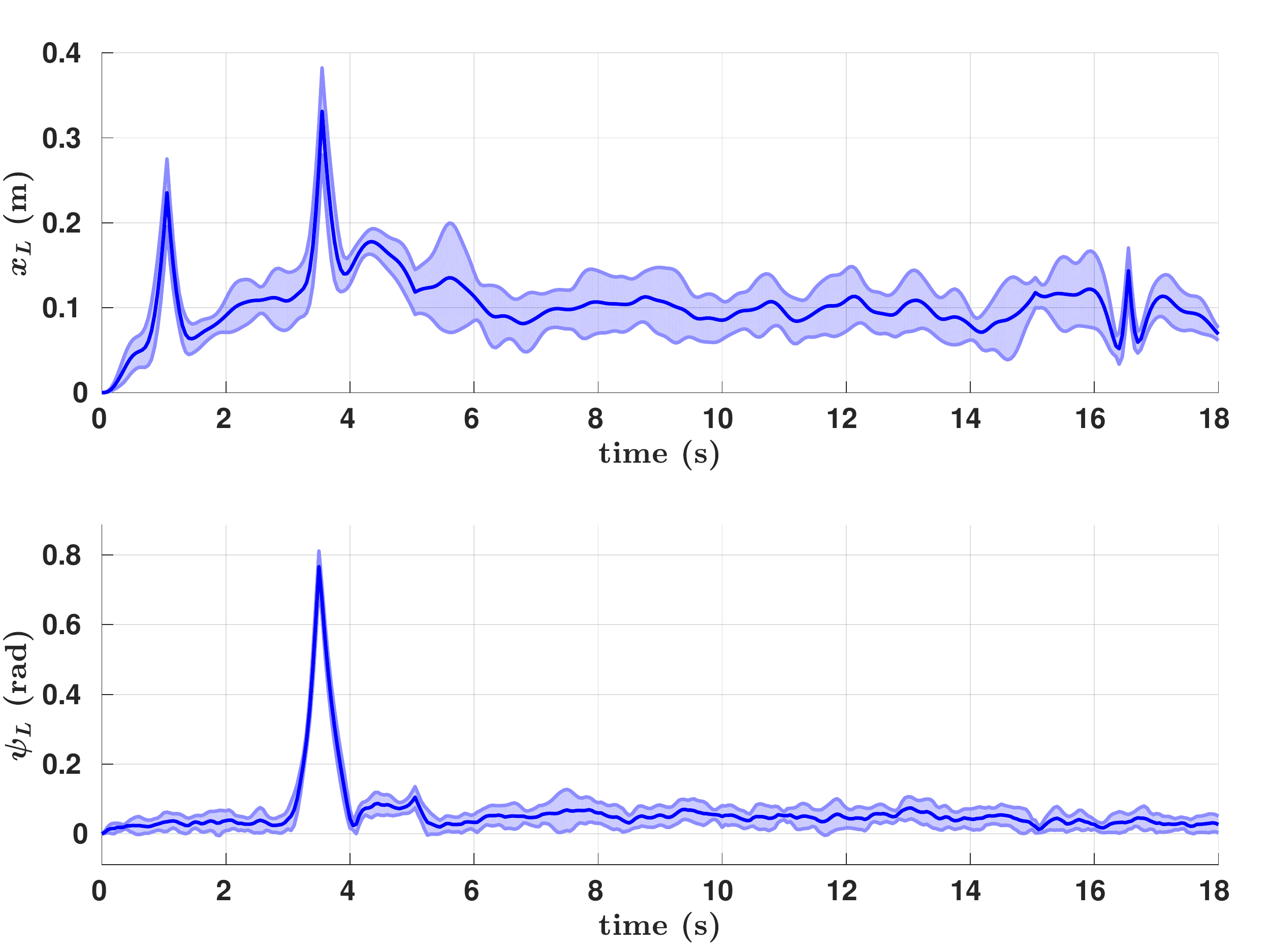}
	\caption {The mean and standard deviations of error norms of absolute position and yaw angle of the leader.}
	\label{fig:TrackErrs}
\end{figure}

\begin{figure}[!]
	\centering
	\includegraphics[scale=0.28]{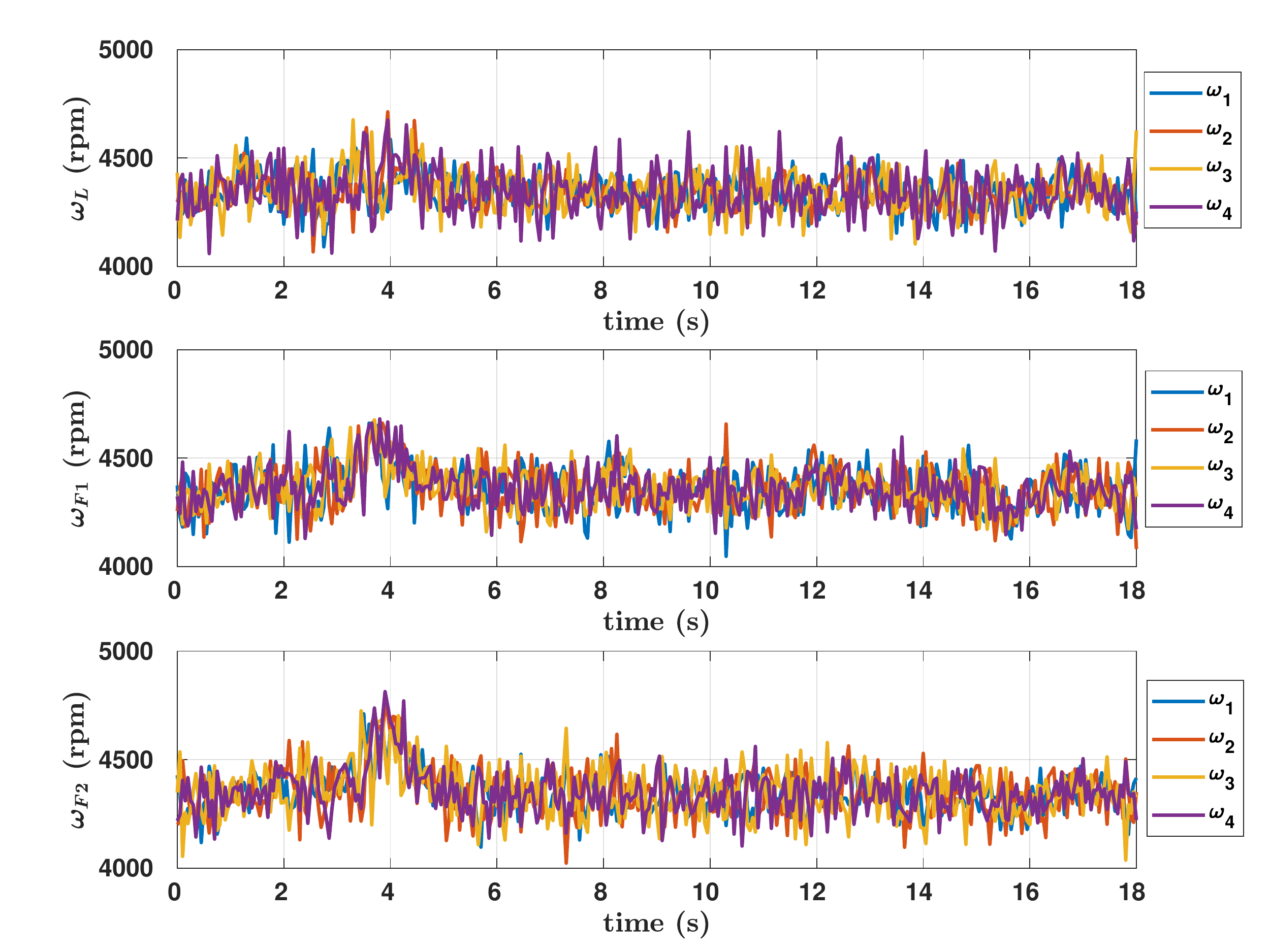}
	\caption {Propeller speeds for the leader and the followers}
	\label{fig:Inputs}
\end{figure}

The error norms of the relative position vectors between the agents are shown in Fig. \ref{fig:FormErrs} and the error norms of the position and the yaw angle of the leader in tracking are illustrated in Fig. \ref{fig:TrackErrs} for 10 runs as well.  As can be observed, the formation error norms are quite low, smaller than 0.068 $ m $, in average and steady-state, and only increases when the scenario changes. The maximum error norm, in steady state, is around 0.015 $ m $, which is also quite low for real applications. Furthermore, these low errors imply that FOV restrictions of relative sensors are also satisfied. However, there are slight absolute position tracking errors in dynamic cases. This is because the formation control weights are much higher than the trajectory tracking. These weights are specified as such since the main focus of this paper is the formation control. Considering the yaw tracking error norms, it is also quite satisfactory. High errors only occurs in the agile maneuvering case which is, again, expected. The propeller speed evolution throughout the scenario for the leader and the followers are illustrated in Fig. \ref{fig:Inputs}. There are two observations that can be made from these results. First, the propeller speeds never exceed the limits and second, although a high frequency chattering is observed in the inputs due to sensor noise, the overall behavior of inputs are quite steady, changes around 600 \textit{rpm} for whole scenario. Finally, the stability of the vehicles can be observed from the submitted video. 

\section{Conclusions}
\label{sec:Conc}
This paper presents a novel problem formulation for controlling formation of the multirotor MAVs. In contrast to existing approaches, the focus is made on the intersection of relative sensing, local coordinates, MPC, 3D space and real applications. After presenting detailed explanation of the problem and the theory, the results of the simulation consisting of different scenarios are illustrated and discussed. The evaluation of results reveals that the satisfactory performance is achieved and maintained both for the formation and the target tracking under model uncertainty and noise.

As future works, the scalability of the method, the communication issues, robust system behavior against non-nominal noise will be investigated. Finally, the decentralized and distributed methods will be studied with the inclusion of state estimators and compared with the current approach.



\bibliographystyle{IEEEtran} 
\bibliography{IEEEabrv,root}

\end{document}